\documentclass[aps,prb,twocolumn,amsmath,amssymb,superscriptaddress,floatfix]{revtex4}
\usepackage{graphicx}
\usepackage{bm}
\usepackage[usenames]{color}
\bibstyle{apsrev.bib}

\newcommand{\be}{\begin{equation}}
\newcommand{\ee}{\end{equation}}
\newcommand{\beqn}{\begin{eqnarray}}
\newcommand{\eeqn}{\end{eqnarray}}

\begin{document}

\title{Entanglement entropy of the $Q \ge 4$ quantum Potts chain}

\author{P\'eter Lajk\'o}
\email{peter.lajko@ku.edu.kw}
\affiliation{Department of Physics, Kuwait University, P.O. Box 5969, Safat 13060, Kuwait}
\author{Ferenc Igl\'oi}
\email{igloi.ferenc@wigner.mta.hu}
\affiliation{Wigner Research Centre, Institute for Solid State Physics and Optics,
H-1525 Budapest, P.O.Box 49, Hungary}
\affiliation{Institute of Theoretical Physics,
Szeged University, H-6720 Szeged, Hungary}
\date{\today}


\begin{abstract}
The entanglement entropy, ${\cal S}$, is an indicator of quantum correlations in the ground state of a many body quantum system. At a second-order
quantum phase-transition point in one dimension ${\cal S}$ generally has a logarithmic singularity. Here we consider quantum spin chains with a
first-order quantum phase transition, the prototype being the $Q$-state quantum Potts chain for $Q>4$ and calculate ${\cal S}$ across
the transition point. According to numerical, density matrix renormalization group results at the first-order quantum phase transition
point ${\cal S}$ shows a jump, which is expected to vanish for $Q \to 4^+$.
This jump is calculated in leading order as $\Delta {\cal S}=\ln Q[1-4/Q-2/(Q \ln Q)+{\cal O}(1/Q^2)]$.
\end{abstract}

\maketitle
\section{Introduction}
\label{sec:intr}
Entanglement is a peculiar feature of quantum mechanics, which is related to the presence of nonlocal quantum correlations.
In a quantum many-body system the entanglement between a spatially confined
region $\mathcal{A}$ and its complement $\mathcal{B}$ is quantified by the entropy\cite{amico,entanglement_review,area}. If the complete
system is in a pure quantum
state $|\psi\rangle$, with a density matrix $\rho=| \psi\rangle\langle\psi |$ then
the entanglement entropy is just the von Neumann entropy of either subsystem
given by
\be
S_{\cal A}=-\bf{Tr} (\rho_{\cal A} \ln \rho_{\cal A}) = 
           -\bf{Tr} (\rho_{\cal B} \ln \rho_{\cal B}) =
S_{\cal B}.
\label{S}
\ee
Here the reduced density matrix is $\rho_{\cal A}=\bf{Tr}_{\cal
B}\rho$, and analogously, $\rho_{\cal B}=\bf{Tr}_{\cal A}\rho$.

The entanglement entropy is a sensitive indicator of quantum correlations in the ground state, therefore
it is used to monitor the different phases and to locate the position of quantum phase transitions.
Most of the studies in this respect are performed in one-dimensional and quasi-one-dimensional objects, such as
in quantum spin chains and ladders. If the total length of the chain is $L$ and the linear size of the subsystem is $\ell$,
then in gapped phases the entanglement entropy has a finite limiting value, as $L \to \infty$ and (then) $\ell \to \infty$.
This observation is a special form of the so called area-law, which states that $S_{\cal A}$ is proportional to the
area of the interface separating $\mathcal{A}$ from the environment\cite{amico,entanglement_review,area}. At a quantum critical point, however, at least in
one dimension the area-law is violated. As the quantum control-parameter, $h$ approaches its critical value, $h_c$,
the characteristic length-scale of quantum fluctuations is divergent: $\xi \sim |h-h_c|^{-\nu}$,
and the entanglement entropy is divergent, too. According to conformal field theory\cite{holzhey,Calabrese_Cardy04} at the critical point $S_{\cal A}(\ell)$
grows logarithmically with $\ell$:
\be
S_{\cal A}(\ell) \simeq \frac{c}{6}b \ln \ell\;,
\label{c}
\ee
where $c$ is the central charge of the conformal algebra and $b$ is the number of contact points between
$\mathcal{A}$ and $\mathcal{B}$: it is $b=2(1)$ for periodic (free) boundary conditions. In the vicinity of the quantum critical
point, when $\ell \gg \xi$ in Eq.(\ref{c}) $\ell$ is replaced by $\xi$. This relation has been verified
analytically and numerically for a set of models\cite{vidal,peschel03,jin_korepin,peschel04,IJ07}. Remarkably, the logarithmic scaling law of entanglement entropy in the
thermodynamic limit is valid even for critical quantum chains that are not conformally
invariant. In those cases the central charge determining the prefactor of the logarithmic
scaling law is replaced by an effective one. Here we mention quantum spin chains with random\cite{refael,Santachiara,Bonesteel,s=1,Laflo05,IgloiLin08,dyn06}  or aperiodic interactions\cite{IJZ07}. On the
other hand in higher dimensions at quantum critical points the logarithmic singularity of the entanglement entropy is generally lost.

In other class of quantum spin chains the quantum phase-transition is first order, such that the derivative
of the ground-state energy density, $\partial e_0(h)/\partial h$ is discontinuous at the transition point. In this way we define (a quantity
analogous to) the latent heat:
\be
\Delta e= \lim_{h \to h_c^+} \frac{\partial e_0(h)}{\partial h} -\lim_{h \to h_c^-}\frac{\partial e_0(h)}{\partial h} \;,
\label{Delta_e}
\ee
and $\Delta e>0$. Similarly the order-parameter (magnetization) is discontinuous, too: it is $m(h)=0$ for $h<h_c$ and
\be
\Delta m= \lim_{h \to h_c^+} m(h)>0\;.
\ee
At the same time the correlation length stays finite at a first-order (quantum) phase-transition point.
It is a basic question how does the entanglement entropy behave at a first-order quantum phase-transition point. It is expected
that in the thermodynamic limit $S_{\cal A}(h)$ shows some kind of singularity as $h$ passes $h_c$.

In this paper we consider a prototype model of first-order quantum phase transitions: the $Q$-state quantum Potts chain\cite{wu}. According to
exact results\cite{baxterbook,baxterpotts,baxtermagn} this model has a first-order transition for $Q>4$.
Here we calculate the entanglement entropy numerically by different methods. For
not too large $Q$ values ($Q=4,6$ and $Q=8$) we use the density matrix renormalization group (DMRG) method\cite{whitePRL}, while in the large-$Q$ limit
a $1/Q$-expansion is performed in leading order.

The structure of the rest of the paper is the following. The quantum Potts model is presented in Sec.\ref{sec:Potts}. The DMRG method and the calculated ground state energy, as well as the latent heat is shown in Sec.\ref{sec:DMRG1}. Results about the entanglement entropy
obtained by the DMRG method and by the $1/Q$-expansion are in Sec.\ref{sec:entanglement}
and discussed in Sec.\ref{sec:disc}. Some details of the calculations on finite chains are given in the Appendix.

\section{Quantum Potts model}
\label{sec:Potts}

The quantum Potts chain is defined by the Hamiltonian\cite{solyompfeuty,igloisolyom1,igloisolyom2}:
\be
{\cal H}=-J\sum_{i=1}^{L} \delta(s_i,s_{i+1}) - h\sum_{i=1}^{L} \sum_{k=1}^{q-1} {\cal M}_i^k\;,
\label{hamilton}
\ee
with $s_i=1,2,\dots,Q$ being a $Q$-state spin variable and ${\cal M}_i$ is a spin-flip operator:
${\cal M}_i^k |s_i\rangle = |s_i+k, {\rm mod}~Q\rangle$. Here we use either periodic chains, when $s_{L+1} \equiv s_1$,
or open chains, when the first sum in Eq.(\ref{hamilton}) runs up to $L-1$.
Often it is more convenient to use another representation
of the model in which the transverse field is diagonal:
\be
{\cal H'}=-\dfrac{J}{Q}\sum_{i=1}^{L} \sum_{k=1}^{Q-1} {\cal M}_i^k {\cal M}_{i+1}^{Q-k}- h\sum_{i=1}^{L} {\cal R}_i\;.
\label{hamilton'}
\ee
(In this representation the states are denoted by $|s'_i\rangle$.)
Here the diagonal elements of the ${\cal R}_i$ operator are $\langle s'_i | {\cal R}_i | s'_i \rangle=-1 + Q \delta(s'_i,1)$.
We note that the definitions of the Hamiltonians in Eqs.(\ref{hamilton}) and (\ref{hamilton'}) can be easily generalized to higher dimensions.

The one-dimensional model is self-dual, its self-duality point is located at $h_c=J/Q$. According to exact results
in the thermodynamic limit ($L \to \infty$) at $h_c$ there is a quantum phase-transition of the system, which is second order for $Q \le 4$
and first order for $Q>4$.
The system is in the ferromagnetic (paramagnetic) phase for $h<h_c$
($h>h_c$). In the second-order regime the critical exponents are known through Coulomb-gas mapping\cite{conjectured} and through conformal invariance\cite{dotsenko,cardy,cardy++}.

In the first-order regime the latent heat is given by\cite{hamer}:
\be
\Delta e=2 \sinh 2 \mu \prod_{n=1}^{\infty} \tanh^2 n \mu\;,
\label{Delta_e}
\ee
in terms of $\mu=\rm{arcosh}(Q^{1/2}/2)$. Close to $Q=4$ the latent heat has an essential singularity:
$\Delta e\sim 4 \pi Q^{1/2} \exp[-\pi^2/2 (Q-4)^{-1/2}]$.
Similarly, the correlation length (inverse mass gap) is expressed as\cite{hamer}:
\be
\xi=\frac{\left[1+2 \sum_{n=1}^{\infty} Q^{n^2}\right]}{2 \sinh 2 \mu}\prod_{n=1}^{\infty} \tanh^{-4} n \mu\;,
\label{xi}
\ee
which behaves close to $Q=4$ as $\xi \sim 1/(8 \pi Q^{1/2}) \exp[\pi^2 (Q-4)^{-1/2}]$.

\section{The DMRG method and results at the transition point}
\label{sec:DMRG1}

The quantum phase-transition in the quantum Potts chain at zero temperature is isomorphic with the temperature driven phase-transition in
the classical two-dimensional model. Using a variant of the DMRG method the properties of the phase-transition in the classical model has been studied both in the second-order\cite{carlonigloi} and in the first-order regime\cite{igloicarlon}. Here the original version of the infinite-size DMRG scheme was utilized for open chains\cite{whitePRL,whitePRB}. The accuracy of the ground state energy calculations was in the range of $10^{-8}-10^{-10}$  and this was in full agreement with the truncation error, the largest basis size being $m=200-300$
for the different systems. Our aim here is to demonstrate the accuracy of the numerical method, which will be then used in Sec.\ref{sec:DMRG2} to study the entanglement entropy of the same model. We concentrate on the $Q=4$ and $Q=6$ models. The first model, being at the border of the second-order transition regime has strong logarithmic corrections\cite{logcorr} and therefore one needs large systems to recover the predicted asymptotic behaviour.

We start to calculate $E_0(L)$, the ground-state energy in a finite system of length $L$ at the phase-transition point with open boundary conditions. At a second-order phase-transition point according to finite-size scaling and conformal invariance the ground-state energy asymptotically behaves as:
\be
E_0(L)=e_0 L + e_1 + e_2(L) +\dots\;,
\label{E_0}
\ee
where $e_0$ and $e_1$ are the bulk energy-density and the surface energy, respectively. The finite-size correction term is universal\cite{bloete,affleck}:
$e_2(L)=-(\pi c v)/(24 L)$, where $v$ denotes the sound velocity and $c$ is the central charge of the Virasoro algebra, which for the $Q=4$ model are given by $v=\pi$ and $c=1$\cite{dotsenko,cardy}.

\begin{figure}[h!]
\begin{center}
\includegraphics[width=9.cm,angle=0]{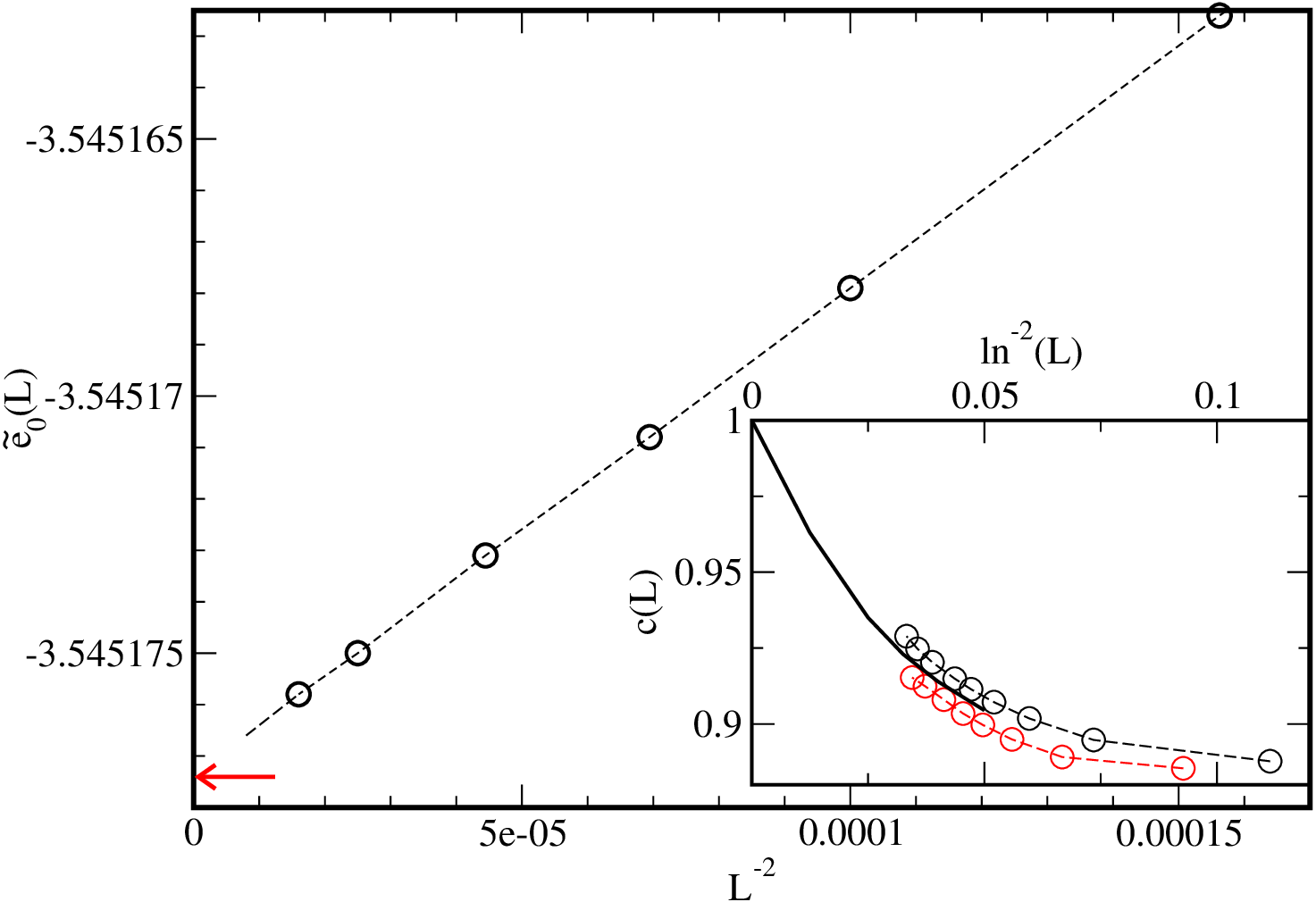}
\caption{\label{fig1} (Color online) Ground-state energy per site of the $Q=4$-state model calculated by DMRG in finite chains, ${\tilde e}_0(L)$ (see text), as a function of $L^{-2}$. The arrow indicates the exact value. In the inset the effective central charge is shown versus $\ln^{-2}(L)$. Upper points: calculated with the use of the exact asymptotic value $e_0=4 \ln 4 -2$; lower points: calculated from two-point fits. The dashed and full lines are guide to the eye.}

\end{center}
\end{figure}

In order to get rid of the surface energy contribution we have calculated the bulk energy-density from the difference: ${\tilde e}_0(L)=[E_0(L+1)-E_0(L-1)]/2$, which has a finite-size correction term: $+(\pi c v)/(24 L^2)$. Indeed, as shown in Fig.\ref{fig1} the DMRG estimates for ${\tilde e}_0(L)$ approach the exact value\cite{hamer}: $e_0=4 \ln 4 -2$ within eight digit accuracy, furthermore the finite-size corrections are quadratic in $1/L$. From the prefactor of the correction term we estimate an effective, size-dependent central charge in the range $0.88 - 0.94$, for increasing $L$, see the inset of Fig.\ref{fig1}. The variation of $c(L)$ with the size is extremely slow, which is due to strong logarithmic corrections\cite{logcorr}. Our data are consistent with the asymptotic form: $c-c(L) \sim \ln^{-2}(L)$, (see also in Ref.\cite{carlonigloi}) and we estimate $c=1.02(6)$ in agreement with the known value $c=1$.

%

\begin{figure}[h!]
\begin{center}
\includegraphics[width=9.cm,angle=0]{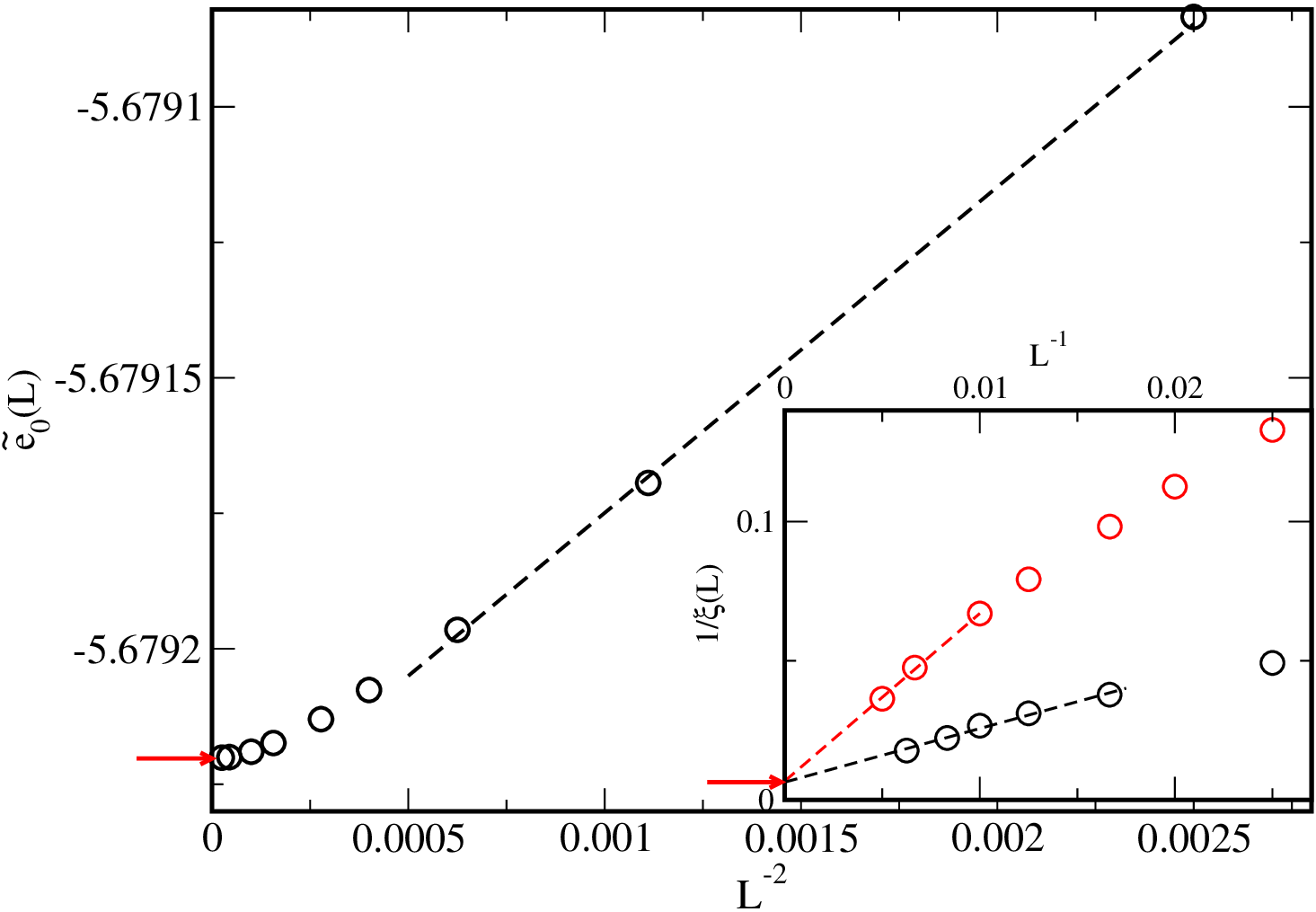}
\caption{\label{fig2} (Color online) Ground-state energy per site of the $Q=6$-state model calculated by DMRG in finite chains , ${\tilde e}_0(L)$ (see text), as a function of $L^{-2}$. The arrow indicates the exact value. In the inset the effective correlation length calculated through Eq.(\ref{L/xi}) is shown as a function of $1/L$. (Lower points: calculated  by two-point fits with the use of the exact asymptotic value of $e_0$; upper points: calculated from three-point fits not using the exact asymptotic value of $e_0$.) The analytical result in Eq.(\ref{xi}) is indicated by an arrow, the dashed lines are guide to the eye.}
\end{center}
\end{figure}

For the $Q=6$ model, for which the phase transition is of first order, the ground-state energy-density, ${\tilde e}_0(L)$ at the phase-transition point is shown in Fig.\ref{fig2}. In this case the finite-size corrections are in the form $\sim 1/L^2$ up to $L<\xi$, which turns to an exponential:
\be
[{\tilde e}_0(L)-e_0] \sim \exp(-L/\xi)\;,
\label{L/xi}
\ee
for $L>\xi$. Here $\xi$ is the correlation length at the transition point, which is estimated through Eq.(\ref{L/xi}) and plotted in the inset of Fig.\ref{fig2}.

%

\begin{figure}[h!]
\begin{center}
\includegraphics[width=9.cm,angle=0]{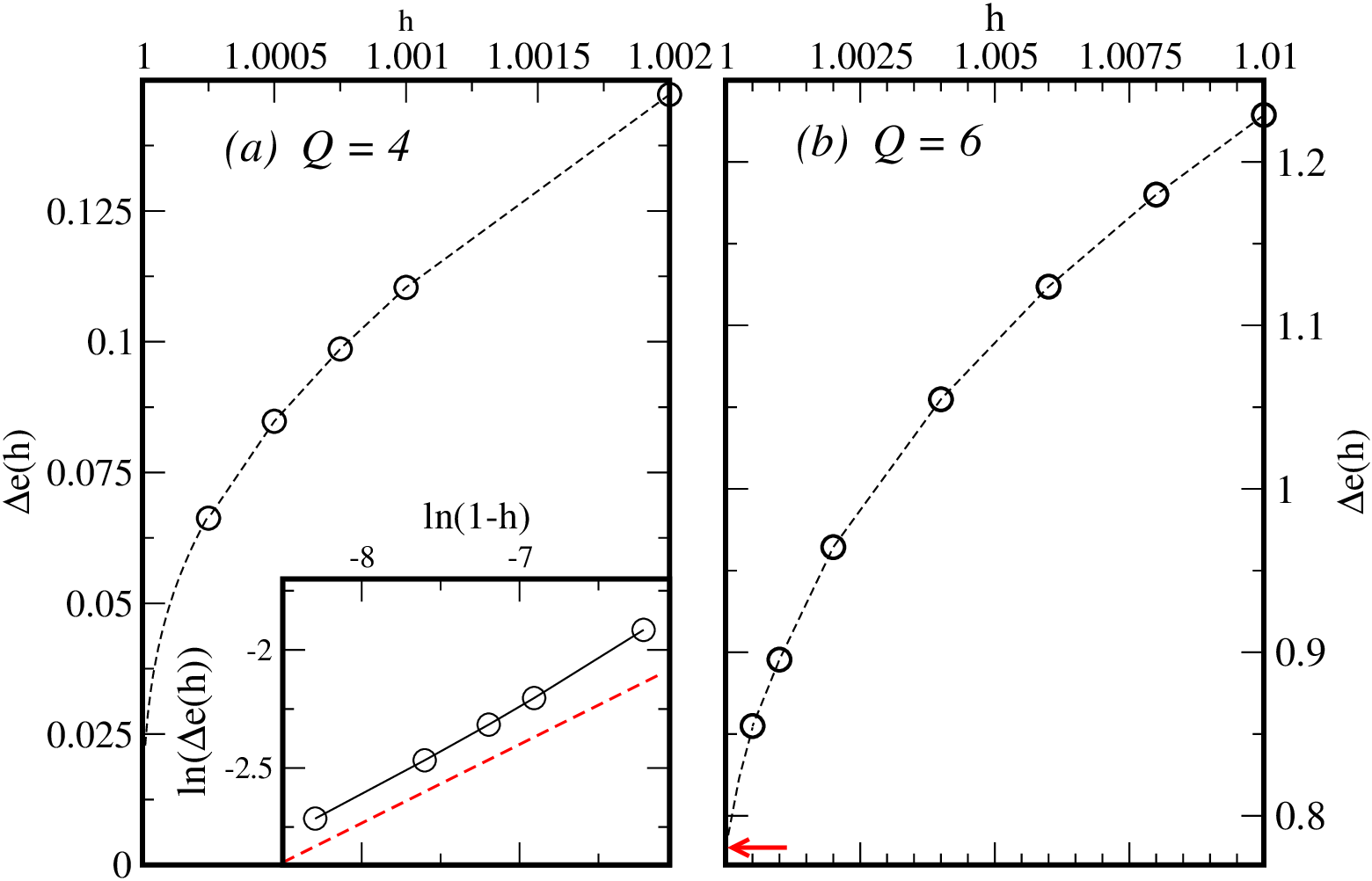}
\caption{\label{fig3} (Color online) Estimates for the effective latent heat as defined in Eq.(\ref{Delta_eh}). Panel a: $Q=4$. In the inset the latent heat vs. (1-h) is shown in log-log plot. The slope of the dashed line is $0.3$. Panel b: $Q=6$. The arrow indicates the analytical result in Eq.(\ref{Delta_e}). The dashed and full lines are guide to the eye.}
\end{center}
\end{figure}

We have also calculated estimates for the latent heat defined by:
\be
\Delta e(h)= \frac{{\tilde e}_0(h)-2{\tilde e}_0(h_c)+{\tilde e}_0(2h_c-h)}{h-h_c} \;,
\label{Delta_eh}
\ee
by the DMRG method at a large value of $L$. Indeed in the limit $L \to \infty$ and $h \to h_c$ $\Delta e(h)$ corresponds to the latent heat in Eq.(\ref{Delta_e}). Estimates for $\Delta e(h)$ are shown in Fig.\ref{fig3} for $Q=4$ (panel a) and for $Q=6$ (panel b). In the first case the effective latent heats tend to zero as $h \to h_c$. As a matter of fact the effective latent heat at a second-order transition has a power-law dependence: $\Delta e(h) \sim (h-h_c)^{1-\alpha}$, where $\alpha$ is the critical exponent of the specific heat, being $\alpha=2/3$ for the $Q=4$ Potts model. This type of scaling form applies for the numerical data, as seen in the inset Fig.\ref{fig3}a. Here in a log-log plot $\Delta e(h)$ vs. $h-1$ is approximately linear and the slope is compatible with $1-\alpha \approx 0.3$. (The relatively slow convergence is due to logarithmic corrections.\cite{logcorr}) On the contrary for the $Q=6$ model the effective latent heats in Fig.\ref{fig3}b tend to a finite limiting value, and $\lim_{h \to 1} \Delta e(h)$ is compatible with the known analytical result in Eq.(\ref{Delta_e}).
%

\section{Entanglement entropy of the Potts chain}
\label{sec:entanglement}

In the calculation of the entanglement entropy we separate the complete (open) system into two halves, such that ${\cal A}$ is
represented by sites $i=1,2,\dots,L/2$ and its complementer ${\cal B}$ consists of $i=L/2+1,L/2+2,\dots,L$. Using the definition
in Eq.(\ref{S}) the entanglement entropy is expressed with the eigenvalues of the reduced density matrix, $\lambda_j$ 
as: ${\cal S}_A= -\sum_j \lambda_j \ln \lambda_j$.

In the two limiting cases, $h=0$ and $h \to \infty$ the entanglement entropy follows from a simple calculation. In the first case, $h=0$,
the ground-state of the system is fully ferromagnetic and given in the representation of Eq.(\ref{hamilton}) as:
\be
|\Psi_0\rangle=\dfrac{1}{\sqrt{Q}}(|11\dots 1\rangle+|22\dots 2\rangle+ \dots +|QQ\dots Q\rangle)\;.
\label{h=0} 
\ee
The reduced density matrix is diagonal and the non-vanishing eigenvalues are:
$\lambda_1=\lambda_2=\dots \lambda_Q=1/Q$, thus the entanglement entropy is: ${\cal S}_A(h=0)=\ln Q$.

In the limit $h \to \infty$ (or $J=0$) the system is fully paramagnetic and it is better to use the transformed basis in (\ref{hamilton'})
in which the ground state is given by:
\be
|\Psi_0\rangle=|1'1'\dots 1'\rangle\;,
\label{J=0} 
\ee
thus the entropy is: ${\cal S}_A(h=\infty)=0$.

For general values of $h$ the entanglement entropy is calculated by different methods. For small finite chains, $L=2$ and $4$ we make analytical calculations, in which $Q$ is a free parameter. For large systems, but for not too large $Q$ values, $Q=4$,$6$ and $8$ we make numerical DMRG calculations. Finally, in the large-$Q$ limit we perform an $1/Q$-expansion in leading order.

\subsection{Solution on finite chains for general values of $Q$}
\label{sec:finite}

In these calculations we calculate first the ground state of the system, for which we use  the transformed basis in Eq.(\ref{hamilton'}).
As explained in the Appendix the eigenstates of ${\cal H'}$ are separated into $Q$ disjoint sectors and the ground-state sector is separated further by symmetry. As a matter of fact the actual ground state is located in the subspace which has the maximal symmetry. This subspace has a finite dimension irrespective of the value of $Q$ and in the solution then $Q$ appears as a (not necessary integer) parameter\cite{solyompfeuty,igloisolyom1,igloisolyom2}. Details of the calculation are explained in the Appendix, where - for simplicity - we choose $J/Q=1$ and with this parametrization the self-duality point is located at $h_c=1$. For $L=2$ and $L=4$ we perform the complete calculation, for larger sizes we consider the large-$Q$ limit, so that our results are correct up to $\mathcal{O}(1/Q)$.

\subsubsection{$L=2$}

The ground state of the system is given by:
\be
|\Psi_0\rangle=a_1 |\phi_1\rangle + \sqrt{1-a_1^2}|\phi_2\rangle\;,
\label{J=0} 
\ee
where $|\phi_1\rangle$ and $|\phi_2\rangle$ are defined in Eq.(\ref{phi_L2}) and
\be
\dfrac{1}{a_1^2}=1+\dfrac{\left[ (h-1)Q/2+1-\sqrt{Q^2(h-1)^2+4Qh}/2\right]^2 }{Q-1}\;.
\label{a_1}
\ee
The reduced density matrix is diagonal having the eigenvalues: $\lambda_1=a_1^2$, $\lambda_2=(1-a_1^2)/(Q-1)$,
$\lambda_3=(1-a_1^2)/(Q-1)$, $\dots$, $\lambda_Q=(1-a_1^2)/(Q-1)$. Thus the entanglement entropy is given by:
\be
{\cal S}_{L=2}=-a_1^2\ln a_1^2 - (1-a_1^2)\ln (1-a_1^2)+(1-a_1^2)\ln (Q-1)\;
\ee
At the critical point, $h=1$, we have $a_1^2=(1+Q^{-1/2})/2$  and the entanglement entropy
for large-$Q$ is given by: ${\cal S}_{L=2}(h=1)\approx \ln Q/2$, which is half of the value measured at $h=0$.

The derivative of the entanglement entropy at the transition point is divergent for large-$Q$: $
\left.\dfrac{\partial {\cal S}_{L=2}}{\partial h}\right|_{h=1} \sim -Q^{1/2}$, and the cross-over regime between
the large- and small-entropy regions has a size: $\delta h_{L=2} \sim Q^{-1/2}$.

\subsubsection{$L=4$}

\begin{figure}[h!]
\begin{center}
\includegraphics[width=8.cm,angle=0]{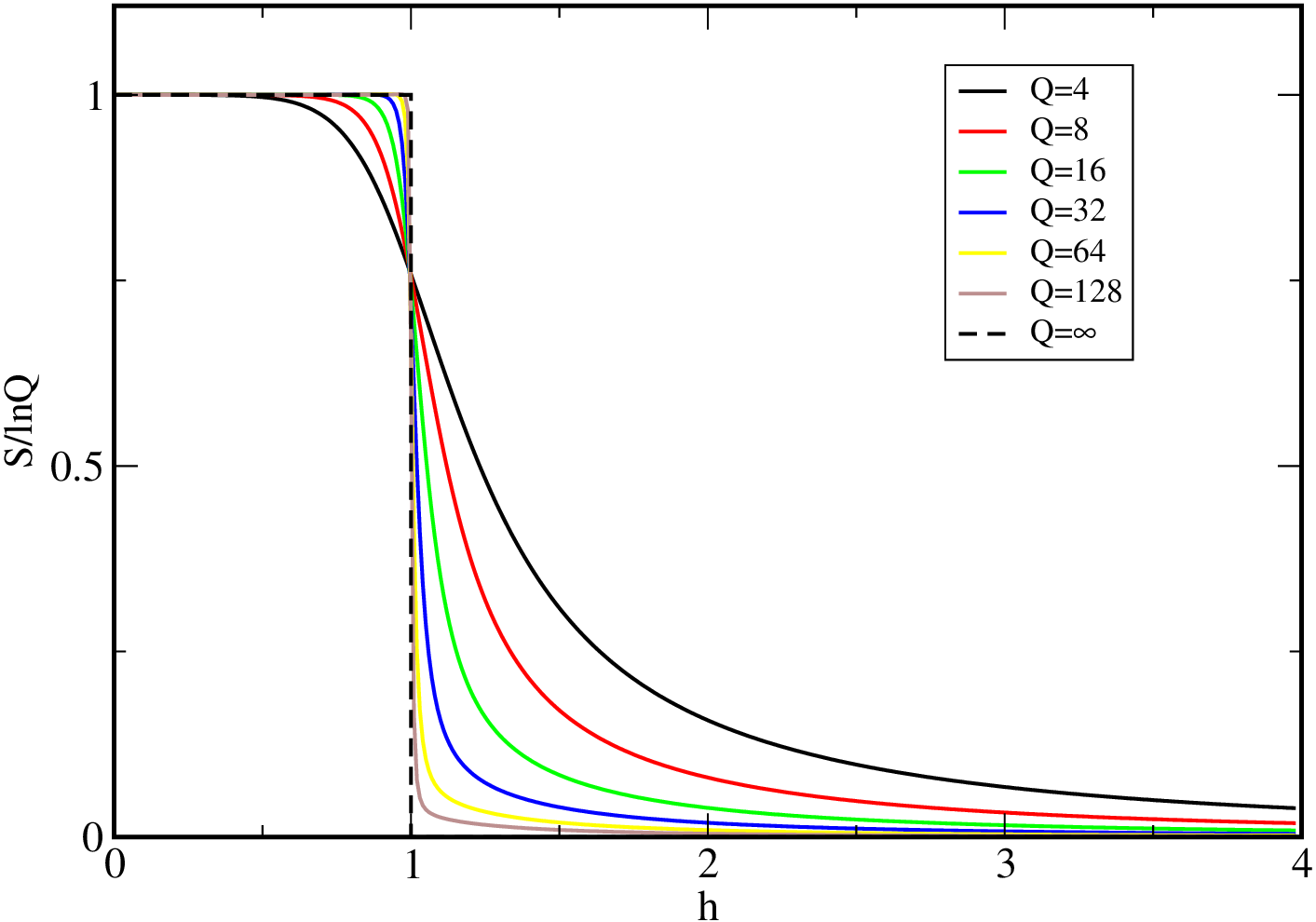}
\caption{\label{fig4} (Color online) Entanglement entropy of the Potts chain with $L=4$ sites for different values of $Q$. }

\end{center}
\end{figure}

The ground-state of the system is given by the linear combination:
\be
|\varPsi_0\rangle=\sum_{i=1}^7 a_i |\phi_i\rangle
\ee
where the basis vectors, $|\phi_i\rangle$, are defined in Eq.(\ref{phi_L4}) and the parameters, $a_i$ are the components of the ground state eigenvector of the matrix in Eq.(\ref{H_L4}). The reduced density matrix, which is of $Q^2 \times Q^2$ is split to $Q$ orthogonal sectors, among which $(Q-1)$-sectors are degenerate. The eigenvalues of the first matrix, $E_i$, $i=1,2,\dots,Q$ are given by Eqs.(\ref{E_1}) and (\ref{E_3}), while the eigenvalues of the second matrix, $\underline{E}_i$, are in Eqs.(\ref{UE_1}), (\ref{UE_2}) and (\ref{UE_3}). The entanglement entropy is expressed in terms of
$\varepsilon_i \equiv E_i \ln E_i$ and $\underline{\varepsilon}_i \equiv \underline{E}_i \ln \underline{E}_i$ as:
\beqn
{\cal S}_{L=4}=&-&\varepsilon_1-\varepsilon_2-(Q-2)\varepsilon_3-\\\;
&-&(Q-1)\left[-\underline{\varepsilon}_1-\underline{\varepsilon}_2-\underline{\varepsilon}_Q 
-(Q-3)\underline{\varepsilon}_3\right]\nonumber
\label{S_L4}
\eeqn
The entanglement entropy of the system as a function of $h$ is shown in Fig.\ref{fig4} for different values of $Q$. Being $L$ finite, the entanglement entropy is analytical function of $h$, but its slope at $h_c=1$ is increasing with $Q$,
asymptotically being $\left.\dfrac{\partial {\cal S}_{L=4}}{\partial h}\right|_{h=1}\sim -Q^{3/2}$. Then the cross-over regime between the large- and small-entropy regions has a size: $\delta h_{L=4} \sim Q^{-3/2}$.
%

\subsubsection{Large-$Q$ limit}
\label{sec:large}

In the large-$Q$ limit the reduced density matrix is analyzed in leading order of $1/Q$ in the Appendix. The correction terms are found to be $L$-independent for $L \ge 4$ and the same holds for the entanglement entropy. This is given by:
\be
{\cal S} \simeq \dfrac{2[\ln \left( Q^2(2h-1)^2\right) +1]}{Q(2h-1)^2},\quad h>1+\delta h, Q \gg 1\;,
\label{S_L4Q1}
\ee
in the disordered phase and
\be
{\cal S} \simeq \ln Q(1+O(1/Q^2)),\quad h<1-\delta h, Q \gg 1\;,
\label{S_L4Q2}
\ee
in the ordered phase. Here the cross-over value of the transverse field, $\delta h$ is $L$-dependent and given in Eq.(\ref{delta_h}).
In the thermodynamic limit $\delta h$ goes to zero, thus the entanglement entropy is discontinuous in the large-$Q$ limit. Its jump at the transition point is given by:
\be
\Delta {\cal S}=\ln Q \left[1-\dfrac{4}{Q} - \dfrac{2}{Q \ln Q}\right] + \mathcal{O}(1/Q^2)\;,
\label{delta_S}
\ee
which is decreasing with decreasing $Q$, and at $Q_c \approx 5.2$
this difference is vanishing. This leading order result is not too far from the exact criterion, $Q_c=4$.

\subsection{DMRG calculations}
\label{sec:DMRG2}
For finite values of $Q$ we have calculated the entanglement entropy numerically by the DMRG method. In these calculations we are focused to the neighbourhood of the transition point, which can not be treated successfully by (large- and small-$h$) expansion methods.

\begin{figure}[h!]
\begin{center}
\includegraphics[width=9.cm,angle=0]{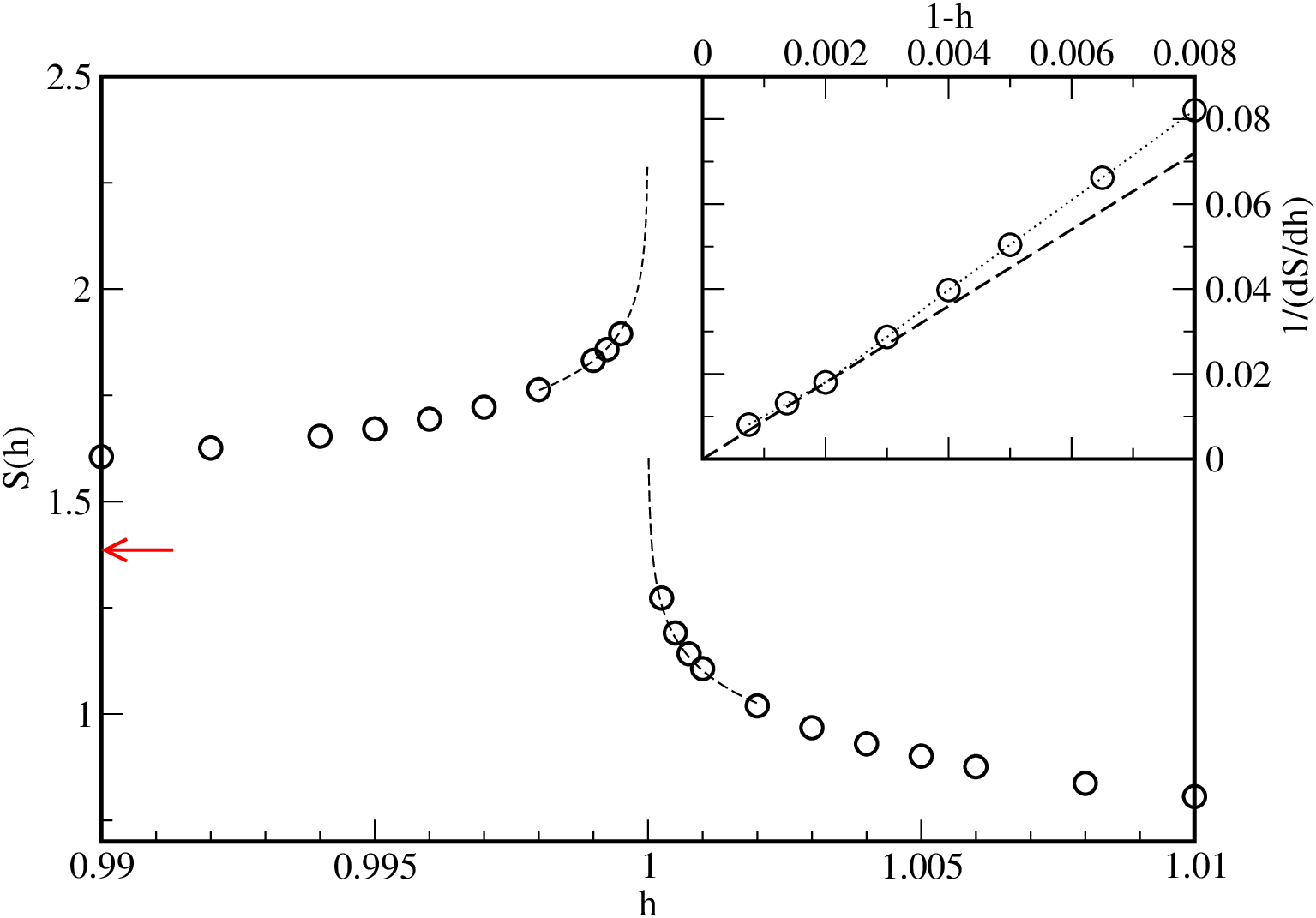}
\caption{\label{fig5} (Color online) Entanglement entropy of the $Q=4$ model in the vicinity of the phase-transition calculated by DMRG. The arrow indicates the limiting value at $h=0$ and the dotted lines are guide to the eye. In the inset the inverse slope is plotted versus $(1-h)$, see Eq.(\ref{Sder}). Here the slope of the dashed straight line is $9$, which corresponds to the analytical prediction.}
\end{center}
\end{figure}

First, in Fig.\ref{fig5} we show results for the $Q=4$ model. In this case the transition is being of second order the entanglement entropy at the critical point is logarithmically divergent, as given in Eq.(\ref{c}). In the vicinity of the transition point in Eq.(\ref{c}) one should replace $\ell$ with the correlation length, $\xi \sim |1-h|^{-\nu}$, thus the slope of the entanglement entropy has a divergence of the form:
\be
\frac{\partial \mathcal{S}}{\partial h} \approx -\frac{c \nu}{6}|1-h|^{-1}\;.
\label{Sder}
\ee
The numerical results in the inset of Fig.\ref{fig5} are in agreement with this prediction, and the prefactor is compatible with the analytical value with $c=1$ and $\nu=2/3$.

%

\begin{figure}[h!]
\begin{center}
\includegraphics[width=9.cm,angle=0]{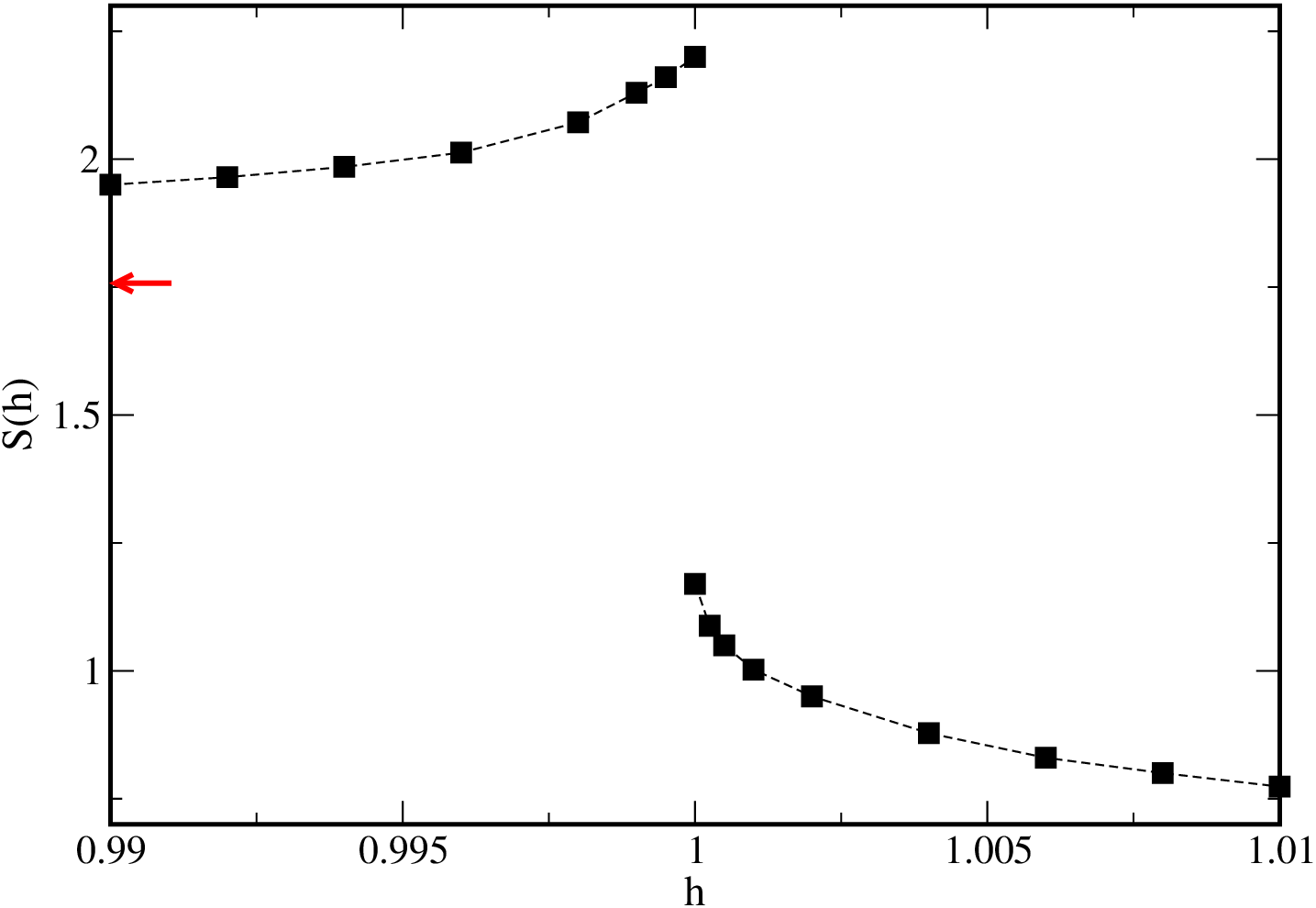}
\caption{\label{fig6} (Color online) Entanglement entropy of the $Q=6$ model in the vicinity of the phase-transition calculated by DMRG.
The arrow indicates the limiting value at $h=0$.}
\end{center}
\end{figure}

Results of a similar calculation for the $Q=6$ model are shown in Fig.\ref{fig6}. In this case the entanglement entropy is monotonously increasing function of $h \le 1$ and the position of its (finite) maximum value is located at the transition point. Then, at $h=1$ it has a jump of $\Delta {\cal S}=1.04(2)$, which is considerably larger than the first-order $1/Q$ result in Eq.(\ref{delta_S}). Thus the higher order terms are quite large for $Q=6$. In the paramagnetic phase the entanglement entropy is monotonously decreasing up to its limiting value $\mathcal{S}=0$, for large $h$.

\begin{figure}[h!]
\begin{center}
\includegraphics[width=9.cm,angle=0]{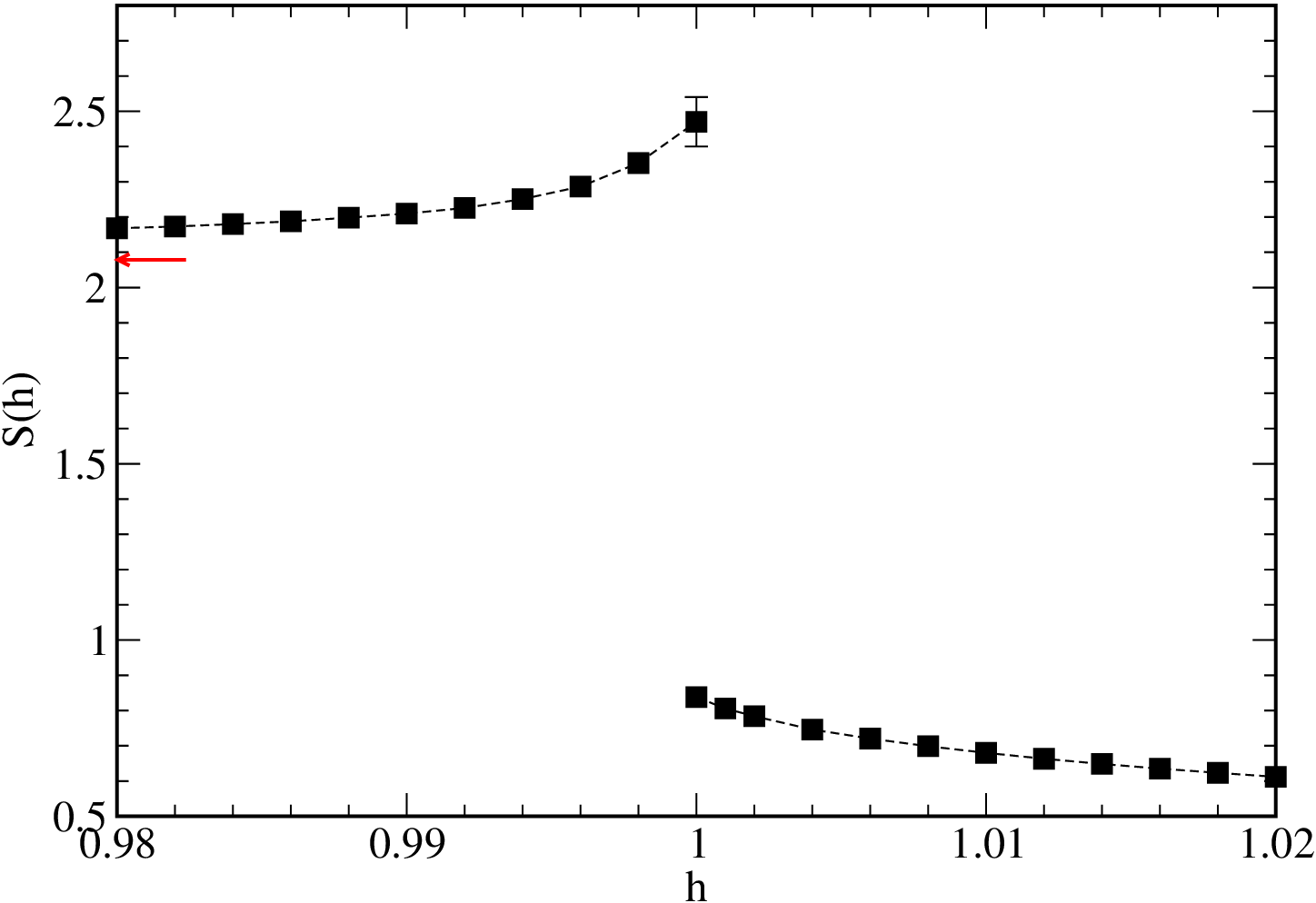}
\caption{\label{fig7} (Color online) The same as in Fig.\ref{fig6} for the $Q=8$ model.}
\end{center}
\end{figure}

We have also studied the $Q=8$ model, in which case the entanglement entropy has similar features as for $Q=6$, see in Fig.\ref{fig7}. In this case the jump of the entropy at the transition point is found $\Delta {\cal S}=1.64(8)$.
%

%

\section{Discussion}
\label{sec:disc}
In this paper we have considered the ferromagnetic quantum Potts chain for $Q \ge 4$ states and studied its entanglement properties close to the phase-transition point. Most of the results are numerical and obtained by the application of the DMRG method. To test the accuracy of the method first we have calculated some known properties of the model (ground-state energy-density, latent heat) and studied their finite-size scaling properties, both for second-order ($Q=4$) and for first-order ($Q=6$) transitions. In this way we have illustrated, how the finite-size scaling behaviour of these quantities at a first-order transition is modified, when the size of the system exceeds the equilibrium correlation length.

The entanglement entropy is found to show different scaling behavior at the transition point, depending on the order of the transition. For $Q=4$ at the second-order transition point the entanglement entropy is logarithmically divergent and the prefactor is observed in agreement with the prediction of conformal invariance. On the other hand for $Q>4$ at the first-order transition point the entanglement entropy stays finite, but develops a finite jump $\Delta {\cal S}(Q)>0$. This jump is a monotonously increasing function of $Q$ and - according to our $1/Q$ expansion - it behaves as $\ln Q$ for large-$Q$. It would be interesting to study by some method the behaviour of $\Delta {\cal S}(Q)$ close to $Q=4$. Here one expects some kind of an essential singularity: $\Delta {\cal S}(Q) \sim \exp[-A (Q-4)^{-1/2}]$, like in the latent heat.
 
To close our paper we mention a few related problems. For \textit{disordered} chains the transition is of second-order, which is controlled by a so-called infinite disorder fixed point\cite{danielreview,im}. Then the entanglement entropy at the critical point follows the logarithmic scaling law in Eq.(\ref{c}) with a so-called effective central charge given by: $c_{\rm eff}=\ln Q/2$\cite{potts_dis}. In the second problem we consider a \textit{single defect}, which connects the two halves of the system, thus it is between sites $i=L/2$ and $i=L/2+1$ and given by $\kappa J$. For the quantum Ising chain with $Q=2$ this problem has already been studied in different papers\cite{peschel03,PeschelZhao,Levine08,ISzL09,Eisler_Peschel10,CMV11,peschel12} and $\kappa$-dependent effective central charge has been obtained. This result is in agreement with the fact, that the defect represents a \textit{marginal perturbation}\cite{ipt} for the Ising chain and the local critical behaviour is $\kappa$-dependent. In the second-order transition regime and for $2<Q \le 4$ the defect is a \textit{relevant perturbation}\cite{ipt}, so that for $0<\kappa<1$ it renormalizes to a cut and for $\kappa>1$ it stays ordered at the transition point. In both cases the entanglement entropy is expected to have a finite, $L$-independent value (see related studies in Ref.\cite{ISzL09}). In the first-order transition regime for $Q > 4$ the previous reasoning does not hold and separate (numerical) investigations are needed to clarify the behaviour of the entanglement entropy. One further question is about the time-dependence of the entanglement entropy of the quantum Potts chain after a non-equilibrium process, such as a global or a local quantum quench\cite{quench_rev}. In the former case the couplings are  changed uniformly and suddenly at $t=0$ and we are interested in the behaviour of ${\cal S}(t)$ for $t>0$. Based on the quasi-particle picture a linearly increasing entanglement entropy, ${\cal S}(t) \sim t$ is expected asymptotically\cite{CC05}. The problem for local quench, when just the strength of a local coupling, say the coupling connecting the two subsystems is changed is more complicated and one can not use results from conformal invariance\cite{EP07,CC07,stephan_dubail}.

\begin{acknowledgments}
This work was supported and funded by Kuwait University Research Grant No.[SP03/15]. The authors thank to Lo\"{\i}c Turban for cooperation in the early stages of the project.
\end{acknowledgments}

\appendix*
\section{Solution on finite chains}

In these calculations we use  the transformed basis in (\ref{hamilton'}) and make use of the fact that eigenstates
of ${\cal H'}$ are separated into $Q$ disjoint sectors. The ground-state sector is characterised by the state:
\be
|\phi_1\rangle = |1'1'\dots 1'\rangle
\ee
and the other $Q^{L-1}-1$ states of the sector are obtained by acting $ {\cal H'}$ on $|\phi_1\rangle$. The
other sectors are characterised with the states: $|2'1'\dots 1'\rangle \rangle$, $|3'1'\dots 1'\rangle \rangle$,
$\dots$,$|Q'1'\dots 1'\rangle \rangle$, and these sectors are degenerate by symmetry. In the following we
concentrate on the ground-state sector and determine its lowest state, which is thus the ground state of the system.
Generally the ground state has the maximal symmetry, which helps us to construct it in a smaller basis set.

\subsubsection{$L=2$}

The ground state of the problem is in the sector $|1'1'\rangle$ having the
maximal symmetry. This subspace is spanned by the vectors:
\beqn
|\phi_1\rangle &=& |1'1' \rangle  \label{phi_L2} \\ \;
|\phi_2\rangle &=&\dfrac{1}{\sqrt{Q-1}}\left[ |2'Q' \rangle + |3'(Q-1)' \rangle + \dots +|Q'2' \rangle\right] \nonumber
\eeqn
We note that for $L=2$ - having periodic boundary conditions - there are two couplings between the two spins.
Then the eigenvalue matrix corresponding to the symmetric subspace is given by:
\be
H_{\rm sym}^{L=2}=\left( \begin{array}{cc}
-2(Q-1)h & -2\sqrt{Q-1} \\
-2\sqrt{Q-1} & 2h-2(Q-2)\end{array} \right)
\ee
and the eigenvector corresponding to the ground state has the components: $a_1$ and $a_2=\sqrt{1-a_1^2}$, which is given in Eq.(\ref{a_1}).

\subsubsection{$L=4$}

In this case the symmetric subspace of the ground-state sector is spanned by the vectors:
\begin{widetext}
\beqn
|\phi_1\rangle &=& |1'1'1'1' \rangle \nonumber \\\;
|\phi_2\rangle &=&\dfrac{1}{\sqrt{4Q_1}}\left[ |1'1'2'Q' \rangle + |1'1'3'(Q-1)' \rangle + \dots +|1'1'Q'2' \rangle +c.p.\right]
\nonumber \\\;
|\phi_3\rangle &=&\dfrac{1}{\sqrt{2Q_1}}\left[ |1'2'1'Q' \rangle + |1'3'1'(Q-1)' \rangle + \dots +|1'Q'1'2' \rangle +c.p.\right]
\nonumber \\\;
|\phi_4\rangle &=&\dfrac{1}{\sqrt{Q_1}}\left[ |2'Q'2'Q' \rangle + |3'(Q-1)'3'(Q-1)' \rangle + \dots +|Q'2'Q'2' \rangle\right] 
\nonumber \\\;
|\phi_5\rangle &=&\dfrac{1}{\sqrt{4Q_1Q_2}}\left[ |1'2'2'(Q-1)' \rangle + |1'2'3'(Q-2)' \rangle + \dots +|1'2'(Q-1)'2' \rangle \right. \nonumber \\\;
&+& \left.|1'3'2'(Q-2)' \rangle + \dots + |1'3'Q'Q' \rangle+ \dots +|1'Q'3'Q' \rangle \right]\nonumber \\\;
|\phi_6\rangle &=&\dfrac{1}{\sqrt{2Q_1Q_2}}\left[|2'Q'3'(Q-1)' \rangle + |2'Q'4'(Q-2)' \rangle  \dots +|2'Q'Q'2' \rangle  
 \right. \nonumber \\\;
&+& \left.  |3'(Q-1)'2'Q' \rangle  + \dots + |3'(Q-1)'Q'2' \rangle+ \dots +|Q'2'(Q-1)'3' \rangle \right]\nonumber \\\;
|\phi_7\rangle &=&\dfrac{1}{\sqrt{Q_1Q_2Q_3}}\left[ |2'2'2'(Q-2)' \rangle + |2'2'3'(Q-3)' \rangle  \dots +|2'2'(Q-2)'2' \rangle  \right. \nonumber \\\;
&+& \left.|2'3'2'(Q-3)' \rangle \dots + |2'3'Q'(Q-1)' \rangle+ \dots +|Q'Q'Q'4' \rangle \right]
\label{phi_L4}
\eeqn
\end{widetext}
and the corresponding eigenvalue matrix is given by:
\begin{widetext}
%
\be
\left( \begin{array}{ccccccc}
-4Q_1h & -\sqrt{4Q_1}       & 0           & 0            & 0          & 0          & 0\\
-\sqrt{4Q_1} & -2Q_2h -Q_2 & -\sqrt{8}  & -2           & -2\sqrt{Q_2}          & -\sqrt{2Q_2}          & 0\\
0           & -\sqrt{8} &  -2hQ_2           & 0            &  -\sqrt{8Q_2}         & 0          & 0\\
0           & -2 &  0           & 4h          & 0  &  -\sqrt{8Q_2}                  & 0\\
0           & -2\sqrt{Q_2} &    -\sqrt{8Q_2}         & 0          & -Q_4h-2Q_2  &  -\sqrt{8}         & -4\sqrt{Q_3}\\
0           & -\sqrt{2Q_2}  & 0 &    -\sqrt{8Q_2}         &  -\sqrt{8}        & 4h-2Q_3  &   -\sqrt{8Q_3}\\
0           & 0  & 0 &    0         &      -4\sqrt{Q_3}   &   -\sqrt{8Q_3} & 4h-4Q_4  

\end{array} \right)
\label{H_L4}
\ee
%
\end{widetext}
Here and in the following we use the short-hand notation, $Q_k \equiv Q-k$, for $k=1,2,3$ and $4$.
The matrix-elemets of the reduced density matrix, $\mathbf{\rho}_A(s'_1s'_2,\tilde{s}'_1\tilde{s}'_2)$, are non-zero only for states
with $s'_1+s'_2-(\tilde{s}'_1+\tilde{s}'_2))=0, {\rm mod}~Q$. The reduced density matrix is devided into $Q$ orthogonal sectors. The first sector contains matrix-elements
between the states: $11,2Q,3(Q-1),\dots,Q2$. In the second sector there are  matrix-elements
between the states: $12,21,3Q,\dots,Q3$. The $n$-th sector is characterized by the states:
$1n,n1,2(n-1),\dots,Q(n+1)$. These latter $(Q-1)$ sectors are degenerate due to symmetry. In the following
we solve the eigenvalue problems of the different sectors.\\
\begin{center}
\textit{$11$ sector}\\
\end{center}
The $Q \times Q$ eigenvalue matrix in this sector is given in the form:

\be
\left( \begin{array}{ccccccc}
a           & c      & c       & c        & c          & \dots        & c\\
c           & b      & d       & d        & d          & \dots        & d\\
c           & d      & b       & d        & d          & \dots        & d\\
c           & d      & d       & b        & d          & \dots        & d\\
c           & d      & d       & d        & b          & \dots        & d\\
\vdots      & \vdots & \vdots  & \vdots   & \vdots     & \ddots       & \vdots\\
c           & d      & d       & d        & d          & \dots        & d  
\end{array} \right)
\label{11sectr}
\ee
with 
\beqn
a &=& a_1^2+a_2^2/4\nonumber \\\;
b &=& \dfrac{1}{4Q_1}(a_2^2+4a_4^2+2a_6^2)\\\;
c &=& \dfrac{a_1 a_2}{2\sqrt{Q_1}}+
\dfrac{a_2 a_4}{2Q_1}+\dfrac{a_2 a_6\sqrt{Q_2}}{\sqrt{8}Q_1}\nonumber \\\;
d &=& \dfrac{a_2^2}{4Q_1}+\dfrac{\sqrt{2}a_4 a_6}{Q_1\sqrt{Q_2}}+\dfrac{a_6^2Q_3}{2Q_1Q_2}\nonumber\;.
\label{abcd}
\eeqn

Two eigenvalues are given in the space of the vectors:
\beqn
\varPsi_1 &=& \varphi_1 \nonumber \\\;
\varPsi_2 &=&\dfrac{1}{\sqrt{Q_1}}\left[ \varphi_2 + \varphi_3 + \dots +\varphi_Q\right]
\eeqn
having an eigenvalue matrix:
\be
\left( \begin{array}{cc}
a & c\sqrt{Q_1} \\
c\sqrt{Q_1} & b+dQ_2\end{array} \right)
\ee
with the eigenvalues:
\be
E_{1,2}=\dfrac{a+b+dQ_2}{2}\pm \sqrt{\left[ \dfrac{a-b-dQ_2}{2}\right]^2 +c^2Q_1 }
\label{E_1}
\ee
The other $(Q-2)$ eigenvalues are degenerate. The corresponding eigenvectors are given in the form:
\beqn
\varPhi_3&=&\dfrac{1}{\sqrt{Q_1}}\left[ \varphi_2 + \eta \varphi_3 + \eta^2 \varphi_3 + \dots +\eta^{Q_2}\varphi_Q\right]
\nonumber \\
 \varPhi_4&=&\dfrac{1}{\sqrt{Q_1}}\left[ \varphi_2 + \eta^2 \varphi_3 + \eta^4 \varphi_3 + \dots +\eta^{2Q_2}\varphi_Q\right]
\nonumber \\
&\vdots& \nonumber \\
 \varPhi_Q&=&\dfrac{1}{\sqrt{Q_1}}\left[ \varphi_2 + \eta^{Q_2} \varphi_3 + \eta^{2Q_2} \varphi_3 + \dots +\eta^{Q_2^2}\varphi_Q\right]
\nonumber \\
\eeqn
with $\eta=\exp\left( \dfrac{2\pi\imath}{Q-1}\right) $. The eigenvalues are:
\be
E_3=E_4=\dots=E_Q=b-d
\label{E_3}
\ee
\\
\begin{center}
\textit{$12$ sector}\\
\end{center}

The $Q \times Q$ eigenvalue matrix in this sector is given in the form:

\be
\left( \begin{array}{ccccccc}
\alpha           & \gamma      & \delta       & \delta        & \delta          & \dots        & \delta\\
\gamma           & \alpha      & \delta       & \delta        & \delta          & \dots        & \delta\\
\delta           & \delta      & \beta       & \varepsilon        & \varepsilon          & \dots        & \varepsilon\\
\delta           & \delta      & \varepsilon       & \beta       & \varepsilon          & \dots        & \varepsilon\\
\delta           & \delta      & \varepsilon       & \varepsilon       & \beta          & \dots        & \varepsilon\\
\vdots           & \vdots & \vdots  & \vdots   & \vdots     & \ddots       & \vdots\\
\delta           & \delta      & \varepsilon       & \varepsilon        & \varepsilon          & \dots        & \beta  
\end{array} \right)
\ee
with
\beqn
\alpha &=& \dfrac{1}{4Q_1}(a_2^2+2a_3^2+a_5^2)\nonumber \\\;
\beta &=& \dfrac{1}{2Q_1Q_2}(a_5^2+a_6^2+2a_7^2) \nonumber \\\;
\gamma &=& \dfrac{a_2 a_3}{\sqrt{2}Q_1}+\dfrac{a_5^2}{4Q_1}\nonumber \\\;
\delta &=& \dfrac{a_2 a_5}{4Q_1\sqrt{Q_2}}
+\dfrac{a_3 a_5}{\sqrt{8}Q_1\sqrt{Q_2}}+\dfrac{a_5 a_6}{\sqrt{8}Q_1Q_2}
+\dfrac{a_5 a_7\sqrt{Q_3}}{2Q_1Q_2}\nonumber \\\;
\varepsilon &=& \dfrac{a_5^2}{2Q_1Q_2}+\dfrac{a_6 a_7\sqrt{2}}{Q_1Q_2\sqrt{Q_3}}+\dfrac{a_7^2Q_4}{Q_1Q_2Q_3}\;.
\label{alpha}
\eeqn
Two eigenvalues are given in the space of the vectors:
\beqn
\underline{\varPsi}_1 &=& \dfrac{1}{\sqrt{2}}\left[\underline{\varphi}_1+\underline{\varphi}_2 \right]\nonumber \\\;
\underline{\varPsi}_2 &=&\dfrac{1}{\sqrt{Q_2}}\left[ \underline{\varphi}_3 + \underline{\varphi}_4 + \dots +\underline{\varphi}_Q\right]
\eeqn
having an eigenvalue matrix:
\be
\left( \begin{array}{cc}
\alpha+\gamma & \delta\sqrt{2Q_2} \\
\delta\sqrt{2Q_2} & \beta+\varepsilon Q_3 \end{array} \right)
\ee
with the eigenvalues:
\beqn
\underline{E}_{1,2}&=&\dfrac{\alpha+\gamma+\beta+\varepsilon Q_3}{2} \label{UE_1} \\ \;
&\pm& \sqrt{\left[ \dfrac{\alpha+\gamma-\beta-\varepsilon Q_3}{2}\right]^2 +\delta^2 2Q_2 } \nonumber
\eeqn
Another $(Q-3)$ eigenvalues are degenerate. The corresponding eigenvectors are given in the form:
\beqn
\underline{\varPhi}_3&=&\dfrac{1}{\sqrt{Q_2}}\left[ \underline{\varphi}_3 + \underline{\eta} \underline{\varphi}_4 + \underline{\eta}^2 \underline{\varphi}_5 + \dots +\underline{\eta}^{Q_3}\underline{\varphi}_Q\right]
\nonumber \\
 \underline{\varPhi}_4&=&\dfrac{1}{\sqrt{Q_2}}\left[ \underline{\varphi}_3 + \eta^2 \underline{\varphi}_4 + \eta^4 \underline{\varphi}_5 + \dots +\eta^{2Q_3}\underline{\varphi}_Q\right]
\nonumber \\
&\dots& \nonumber \\
 \underline{\varPhi}_{Q-1}&=&\dfrac{1}{\sqrt{Q_2}}\left[ \underline{\varphi}_3 + \underline{\eta}^{Q_3} \underline{\varphi}_4 + \underline{\eta}^{2Q_3} \underline{\varphi}_5 + \dots +\underline{\eta}^{Q_3^2}\underline{\varphi}_Q\right]
\nonumber \\
\eeqn
with $\underline{\eta}=\exp\left( \dfrac{2\pi\imath}{Q-2}\right) $. The eigenvalues are:
\be
\underline{E}_3=\underline{E}_4=\dots=\underline{E}_{Q-1}=\beta-\varepsilon
\label{UE_2}
\ee
Finally, last eigenvector is given by:
\be
\underline{\varPhi}_Q=\dfrac{1}{\sqrt{2}}\left[\underline{\varphi}_1-\underline{\varphi}_2 \right]
\ee
with the eigenvalue:
\be
\underline{E}_Q=\alpha-\gamma
\label{UE_3}
\ee
The entanglement entropy of the system is expressed in terms of the eigenvalues, $E_i$ and $\underline{E}_i$ and given in Eq.(\ref{S_L4}).

\subsubsection{Large-$Q$ limit}

For large-$Q$ we consider the leading behaviour up to $1/Q$, when the matrix-elements of the reduced density matrix are different in the disordered and in the ordered phase, respectively. We start with the analysis of the results of the previous subsection for the $L=4$ chain. 

\underline{In the disordered regime, $h>1+\delta h$,} in leading order the following matrix-elements are non-zero:
\beqn
a&=&1-\dfrac{3}{Q(2h-1)^2},\quad b=d=\alpha=\dfrac{1}{Q^2(2h-1)^2}\nonumber \\\;
c&=&\dfrac{1}{Q(2h-1)}\;
\eeqn
and the leading contribution to the entropy is given in Eq.(\ref{S_L4Q1}).

\underline{In the ordered regime, $h < 1-\delta h$,} the non-zero matrix-elements in leading order are the following:
\beqn
\alpha&=&\gamma=\dfrac{4}{Q^2(2-h)^2},\quad \delta=\dfrac{2}{Q^2(2-h)},\\\;
b&=&d=\dfrac{1}{Q^2},\quad \beta=\varepsilon=\dfrac{1}{Q^2}\left[1-\dfrac{1}{Q}\left(\dfrac{8}{(2-h)^2}+1\right)  \right]\nonumber \;
\eeqn
and the leading contribution to the entropy is given in Eq.(\ref{S_L4Q2}).

Finally, \underline{at the phase-transition point, $h=1$,} the matrix-elements in leading order are $a=1/2$, $\beta=1/[(q-1)(q-2)]$ and $\varepsilon=(q-4)/[(q-1)(q-2)(q-3)]$ and the leading contribution to the entropy is given by
\be
{\cal S}_{L=4} \simeq \dfrac{1}{2}\ln (4q)(1+O(1/q)),\quad h=1, q \gg 1\;.
\label{S_L4Q3}
\ee

For general value of $L \ge 4$ we use a perturbation calculation.

\underline{In the large-$h$ limit} the symmetrical subspace is spanned by two verctors:
\beqn
|\phi_1\rangle &=& |1'1'1' \dots 1'1'1' \rangle \nonumber \\\;
|\phi_2\rangle &=& \dfrac{1}{\sqrt{LQ_1}}\left[ |1'1'1'\dots1'2'Q' \rangle + |1'1'1'\dots1'3'(Q-1)' \rangle\nonumber \right. \\\;
 &+&\left. \dots +|1'1'1'\dots1'Q'2' \rangle + c.p. \right]\;
\label{phi_LL}
\eeqn
and the ground state is given by: $|\Psi_0\rangle=a_1 |\phi_1\rangle + a_2|\phi_2\rangle$, with $a_2=a_1\dfrac{\sqrt{L}}{\sqrt{q}(2h-1)}$ and $a_1^2+a_2^2=1$. Another excited states have no contribution, provided $h >1+\delta h$ and 
\be
\delta h \sim q^{-L/4-1/2} \sim \exp(-L/\xi)\;,
\label{delta_h}
\ee
with $\xi=4/\ln q$ being the correlation length for large-$q$.

The reduced density matrix has the same structure, as for $L=4$: there is a non-degenerate subspace and a $(Q-1)$-fold degenerate one. In both subspaces just the largest eigenvalue is of the order of at least $1/Q$ and these are: $E_{1}=1-\dfrac{2a_2^2}{L}=1-\dfrac{2}{Q(2h-1)^2}$ and $\underline{E}_{1}=\dfrac{a_2^2}{L(Q-1)}=\dfrac{1}{Q(Q-1)(2h-1)^2}$. Consequently the leading $1/Q$-correction to the entanglement entropy, for $h>1+\delta h$ is the same for any chain of length $L \ge 4$, and given in Eq.(\ref{S_L4Q1}).

\underline{In the small-$h$ limit} we work in the original basis in Eq.(\ref{hamilton}) and the ground state in leading order is given by: $|\Psi_0\rangle=b_1 |\phi_1\rangle + b_2|\phi_2\rangle$. Here $|\phi_1\rangle$ is given in the same form as in Eq.(\ref{h=0}) and
\be
|\phi_2\rangle = \dfrac{1}{\sqrt{LQ(Q-1)}}\left[ \sum |ii\dots iji\dots i\rangle\right]
\ee
for $i=1,2,\dots,Q$, $j \ne i$ and there are $L$ different positions of the state $|j\rangle$. The weights in the ground state are:
\beqn
b_1=1-h^2\dfrac{L(Q-1)}{2(2-h)^2 Q^2},\quad
b_2=h\dfrac{\sqrt{L(Q-1)}}{Q(2-h)}\;.
\eeqn
The reduced density matrix is split into $Q$ identical sectors, each of which is
characterised by the state $j=1,2,\dots,Q$. The non-vanishing diagonal $(A,B)$ and off-diagonal $(C,D)$ matrix-elements are:
\beqn
A&=&\dfrac{b_1^2}{Q}+\dfrac{b_2^2}{2Q}, \quad B=\dfrac{b_2^2}{LQ(Q-1)},\nonumber\\\;
C&=&\dfrac{b_1}{\sqrt{Q}}\dfrac{b_2}{\sqrt{LQ(Q-1)}}, \quad D=\dfrac{b_2^2}{LQ(Q-1)} \;.
\eeqn
In one of the sectors, the reduced density matrix has the same form as given in Eq.(\ref{11sectr}), with the correspondences:
$a \to A$, $b \to B$, $c \to C$ and $d \to D$, but the dimension of the
matrix is $r=1+(Q-1)L/2$ instead of $Q$. With this the first two eigenvalues follows from Eq.(\ref{E_1}), which in leading order are $E_1=\dfrac{1}{Q}\left[1+\mathcal{O}(Q^{-2}) \right]$ and $E_2=\mathcal{O}(Q^{-3})$, while $E_3=E_4=\dots=E_r=B-D=0$.
Consequently the entanglement entropy in leading order is given by Eq.(\ref{S_L4Q2}).

\end{document}